\documentclass[pre,twocolumn,preprintnumbers,amsmath,amssymb]{revtex4}
\usepackage{graphicx}
\usepackage{dcolumn}
\usepackage{bm}
\usepackage{fullwidth}
\usepackage{float}
\usepackage{placeins}

\begin{document}

\title{Comparative visualization of epidemiological data during various stages of a pandemic}

\author{Thomas Kreuz}
\email{thomas.kreuz@cnr.it}
 \affiliation{Institute for Complex Systems, CNR, Sesto Fiorentino, Italy}

\date{\today}

\begin{abstract}
\noindent After COVID-19 was first reported in China at the end of 2019, it took only a few months for this local crisis to turn into a global pandemic with unprecedented disruptions of everyday life. 
However, at any moment in time the situation in different parts of the world is far from uniform and each country follows its own epidemiological trajectory.
In order to keep track of the course of the pandemic in many different places at the same time, it is vital to develop comparative visualizations that facilitate the recognition of common trends and divergent behaviors.
Similarly, it is important to always focus on the information that is most relevant at any given point in time.
In this study we look at exactly one year of daily numbers of new cases and deaths and present data visualizations that compare many different countries and are adapted to the overall stage of the pandemic.
During the early stage when cases and deaths still rise we focus on the time lag relative to the current epicenter of the pandemic and the doubling times. 
Later we monitor the rise and fall of the daily numbers via wave detection plots.
The transition between these two stages takes place when the daily numbers stop rising for the first time.
\end{abstract}

%\keywords{}

\maketitle

% ####################################################################################
% ####################################################################################
% ############################                          ##############################
% ############################       Introduction       ##############################
% ############################                          ##############################
% ####################################################################################
% ####################################################################################

\section{\label{sec:Intro} Introduction}

%To-do list: Title, Abstract, figure placement, check ZZZZZ, add a few more citations.

\noindent Human cases of COVID-19, the disease caused by the novel coronavirus Severe Acute Respiratory Syndrome Corona-Virus 2 (SARS-CoV-2), were first reported in Wuhan City, China, in December 2019 \citep{huang2020clinical}. After initial transmissions were restricted to Central China’s Hubei province, already by January 2020 first cases had been reported not only in other Asian countries (starting with Taiwan, South Korea and Japan) but also in Australia, the US and several European countries \citep{world2020coronavirus11}. Similarly to what had happened before in China  \citep{zhu2020novel, guan2020clinical}, within February 2020 first cases turned into first deaths \citep{zhou2020clinical} and South Korea, Iran and increasingly Italy emerged as early hotspots outside of China where the epidemic now seemed to be under control \citep{world2020coronavirus40}. On March 11, 2020, the Director General of the World Health Organisation (WHO) declared the novel coronavirus outbreak a worldwide pandemic \citep{world2020coronavirus51}. A few days later Italy became the clear global epicenter as the first country to surpass China in number of deaths \citep{world2020coronavirus60}. Since then the coronavirus has spread across the globe with an unprecedented impact on healthcare \citep{shamasunder2020covid}, economy \citep{mckibbin2020economic}, finances \citep{zhang2020financial}, science \citep{palayew2020pandemic, myers2020unequal}, education \citep{marinoni2020impact}, travel \citep{vskare2020impact}, sports \citep{garcia2020impact}, mental health \citep{dubey2020psychosocial} and basically all other sectors of society.

Already on January 22, 2020, the Center for Systems Science and Engineering (CSSE) at John Hopkins University (Baltimore, MD, USA) started publishing a freely available COVID-19 Data Repository  that was updated daily \citep{dong2020interactive}. Once global datasets like this one became publicly available, the scientific community sprang into action and within a short time a host of studies appeared that focused to a large extent on modeling the data and using these models to predict the future course of the pandemic (e.g., \citep{fanelli2020analysis, flaxman2020estimating, dehning2020inferring, desousa2020kinetic}). 

While modelling the Covid-19 pandemic has already yielded some important results and policy recommendations \citep{sy2020policy, moon2020fighting, jia2020modelling} and much work still remains to be done \citep{vespignani2020modelling, prakash2020minimal}, extrapolation based on often incomplete, inaccurate or unreliable data does also have its limitations and pitfalls \citep{sridhar2020modelling}. In this article we refrain from making inferences about the future but rather restrict ourselves to pure visualization of past and present data \citep{callaghan2020covid}. The aim, in a nutshell, is to develop a simple and consistent comparative data visualization framework that is general and adapted to the various stages of a pandemic, thereby summarizing in one sweep the dynamic and heterogeneous situation worldwide.

First, we focus on visualizations that allow a meaningful comparison of the course of the pandemic for many different countries (or on smaller spatial scales: states, regions etc.) at the same time. This is in contrast to the commonly used histograms in which the temporal profile of the data is plotted for one country at a time. Second, we argue that the most relevant information to be found in the data changes between the early and the later stages of a pandemic and accordingly we present two different kind of data visualizations.

In the early stage (as long as cases and deaths continue to rise) it is most important to monitor both the time lag compared to the current epicenter (typically the country where the epidemic first took hold but later this can shift) and the severity of the spread of the disease (usually expressed by means of the doubling time). Both of these quantities provide very useful information about the urgency of the situation \citep{lancet2020covid} and can help with general decision making in order to find the right moment for the implementation of preventive measures \citep{phua2020intensive, kennedy2020modeling}.

On the other hand, the later stages are more about monitoring the course of the pandemic in each country in terms of peaks, valleys and plateaus. This is when there is often a back and forth between imposing, tightening and relaxing of contact restrictions depending on the trajectory of the epidemiological dynamics in the population at that point in time \citep{goldsztejn2020public, charpentier2020covid}. The transition between these two stages takes place when the daily numbers of cases and deaths stop rising for the first time \citep{leung2020first}.

The remainder of this article is organized as follows: First in Section \ref{s:Data} we describe the dataset and the preprocessing performed. The two Method Sections \ref{ss:Methods-EarlyStages} and \ref{ss:Methods-LaterStages} illustrate the quantities, graphs and sorting criteria we will use to visualize the data in the early and the later stages of the pandemic, respectively. In Sections \ref{ss:Results-EarlyStages} and \ref{ss:Results-LaterStages} we show the data plots for $42$ selected countries from all over the world as well as two more local examples, the US states and the regions of Italy. Finally, in Section \ref{s:Discussion} we summarize and conclude.

% ####################################################################################
% ####################################################################################
% ############################                          ##############################
% ############################           Data           ##############################
% ############################                          ##############################
% ####################################################################################
% ####################################################################################

\section{\label{s:Data} Data}

% Figure 1
\begin{figure*}
	\begin{minipage}{1.07\textwidth}
		\includegraphics[width=1\textwidth]{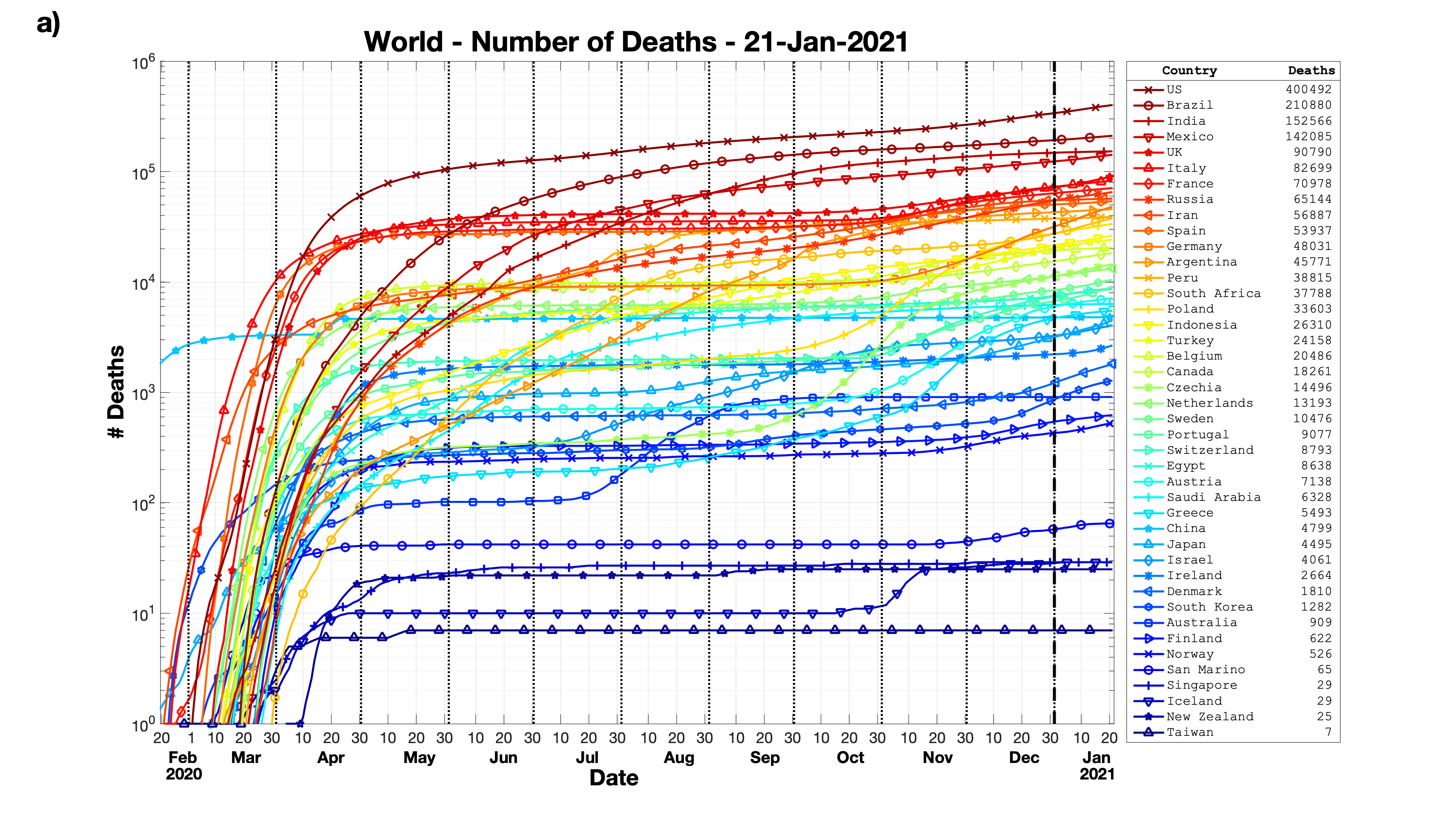}
		\includegraphics[width=1\textwidth]{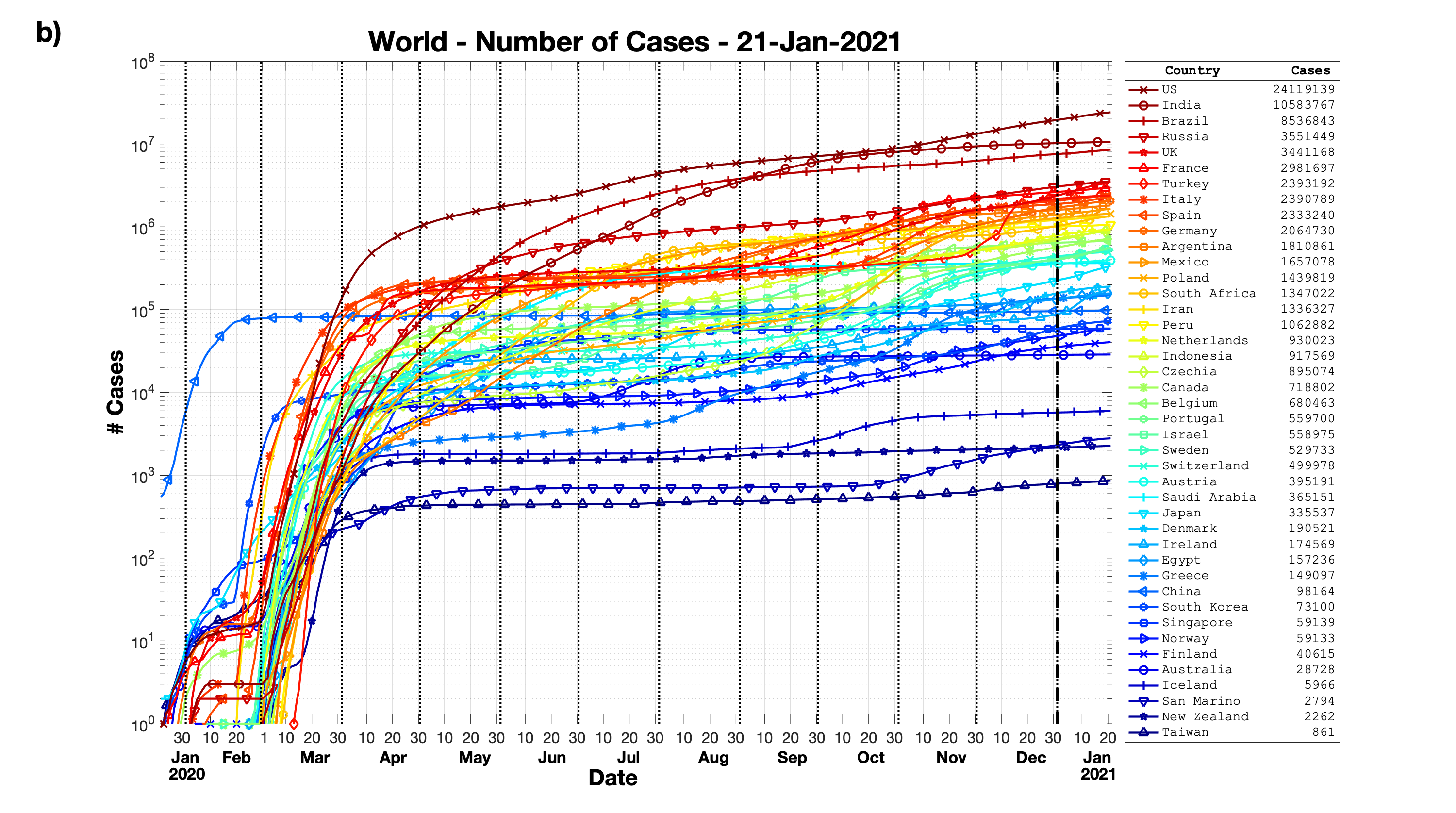}
	\end{minipage}
	\caption{The dataset: Absolute numbers of cumulative deaths (a) and cases (b) versus time for 42 selected countries and the one year interval from January 22, 2020 to January 21, 2021. This plot like all subsequent plots that show cumulative numbers, uses a log scale to facilitate the comparison also between countries at different stages of their pandemic course.	The legend lists the countries and their cumulative number of cases and deaths after one year, respectively. Source: John Hopkins COVID-19 Data Repository.}
	\label{Fig1_Data}
\end{figure*}
\textbf{Dataset:} We illustrate our data visualizations using the freely available dataset from the COVID-19 Data Repository by the Center for Systems Science and Engineering (CSSE) at Johns Hopkins University (Github Webpage: \url{https://github.com/CSSEGISandData/COVID-19}) \citep{dong2020interactive}. We here use the daily cumulative data for both cases and deaths right from its first publication on January 22, 2020, until January 21, 2021, thus covering exactly one year of data. We selected $42$ representative countries focusing on Europe and larger countries in other parts of the world. The data for the US states were available within the same repository. The data for the Italian regions can be found on the webpage \url{https://github.com/pcm-dpc/COVID-19}. The data from Liguria were not available and this region was thus not included. The population sizes used in the normalization were taken from the webpage \url{https://www.worldometers.info/coronavirus/} (data as of January 21st, 2021).

% Preprocessing:
\textbf{Preprocessing:} Datasets of both the overall number of cases and deaths up to a certain date can be expected to increase monotonically with time, but not strictly monotonously since there might be days without any new cases or deaths. However, occasionally some of the datasets do contain negative jumps from one day to the next, typically due to elimination of double counting or other kinds of retrospective reevaluations such as fundamental changes in the way cases and deaths were defined (see the COVID-19 Data Repository \citep{dong2020interactive} for details). 
To clean the data and eliminate these negative jumps we follow the reasonable assumption that later data are more correct than earlier data (after all that is what corrections are for) and each time decrease all the spuriously high early data points to the corrected later value. As a positive side effect this smoothing also eliminates most of the plateaus that were created due to the aforementioned corrections, and this helps in the later calculations of the doubling times. Finally, we apply a moving average of order 7 days in order to smooth out weekday variations such as the tow-day dips that often occur due to reporting delays on weekends \citep{bergman2020oscillations}.

%%%%%%%%%%%%%%%%%%%%%%%%%%%%%%%%%%%%%%%%%%%%%%%%%%
%%%%%%%%%%%%%%%%%%%%% Fig. 1 %%%%%%%%%%%%%%%%%%%%%
%%%%%%%%%%%%%%%%%%%%%%%%%%%%%%%%%%%%%%%%%%%%%%%%%%
%
The cumulative absolute (not normalized by population) number of deaths and cases for the selected $42$ countries and the year from January 22, 2020, to January 21, 2021, obtained in this way are shown in \textbf{Fig. \ref{Fig1_Data}a} and \textbf{Fig. \ref{Fig1_Data}b}, respectively.

%
%\begin{figure*}[!ht]
%	\centering
%	\includegraphics[width=0.4\textwidth]{Corona-Paper-Fig1c_Monotony-Correction.png}
%	\caption{Corona 1c: Corrections. Instead of increasing the later values to the early spurious maximum (red curve) we rather decrease the earlier spurious values themselves.}
%	\label{Fig1c_Corrections}
%\end{figure*}
%
% Please note that I corrected obvious glitches in the data (e.g. some missing updates) using the webpage https://www.worldometers.info/coronavirus/.
%

% ####################################################################################
% ####################################################################################
% ############################                          ##############################
% ############################          Methods         ##############################
% ############################                          ##############################
% ####################################################################################
% ####################################################################################

\section{\label{s:Methods} Methods}

% Figure 2
\begin{figure*}[!ht]
	\centering
	\includegraphics[width=1\textwidth]{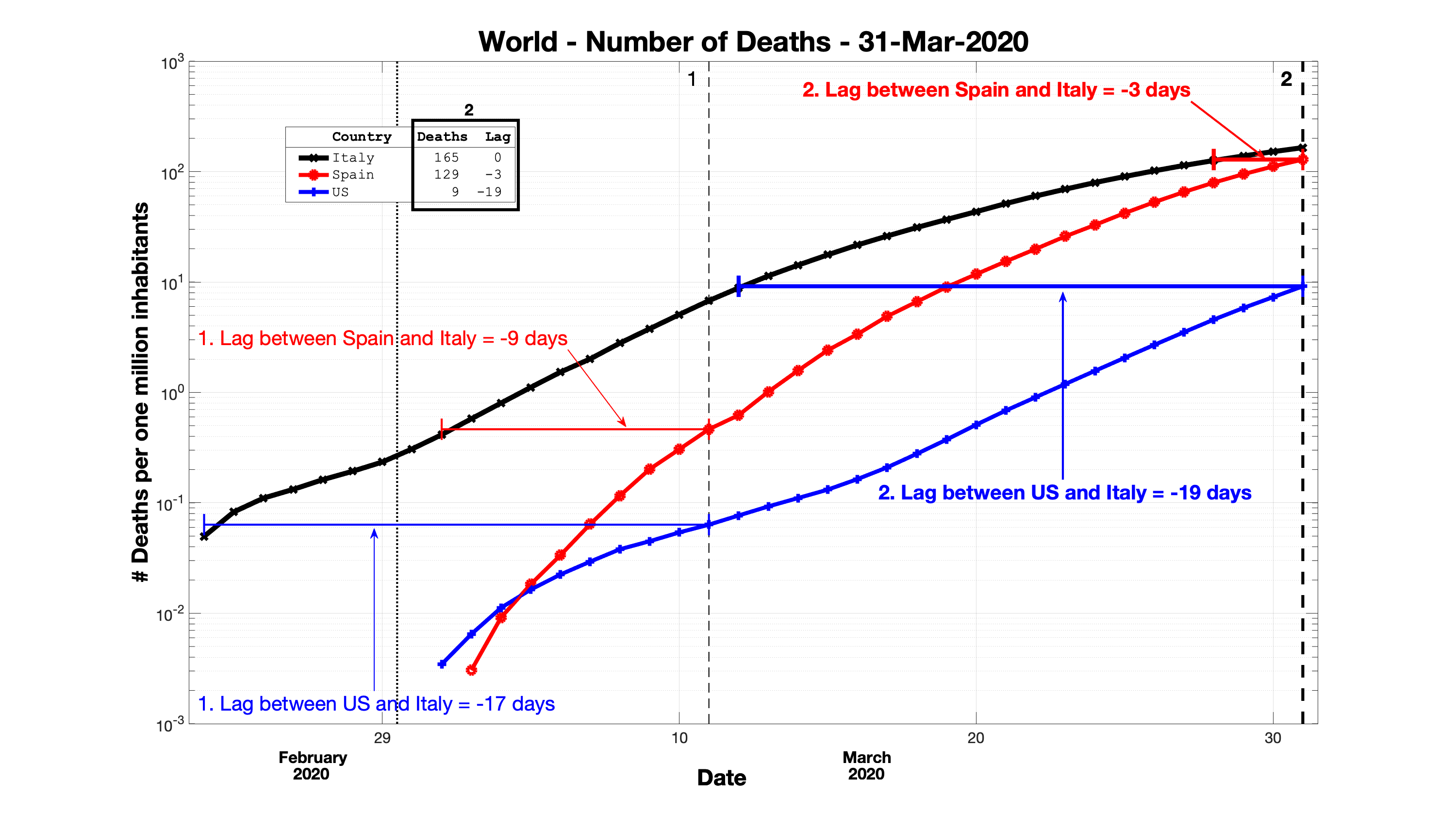}
	\caption{Definition of time lags: The monotonously increasing thick curves depict normalized number of deaths (per one million inhabitants) versus time for three selected countries during the time interval from February 23 to March 30, 2020. Horizontal lines illustrate the changing time lag of Spain (red) and the US (blue) with respect to the pandemic course of Italy (thick black curve), first for March 11, 2020 (label 1 and thin lines on the left) and then for the end of March (label 2 and thick lines on the right). During these twenty days the time lag between Spain and Italy decreased from $-9$ to $-3$ days, while the time lag between the US and Italy increased from $-17$ days to $-19$ days. The time lags on March 31, together with the number of deaths up to that day, are reported in the legend.}
	\label{Fig2_Time_Lags}
\end{figure*}

\noindent In the Method Section we illustrate the various data visualization plots using the number of deaths as an example. Deaths tend to be more reliable \citep{subbaraman2020daily} since they are not affected by the number of tests performed which itself depends on a variety of factors, not only the number of either symptomatic or essential people (compare, e.g., \cite{johansson2021sars, pullano2021underdetection}) but also healthcare system capacities and political decisions \citep{callaghan2020covid}. However, for completeness, in the Results Section \ref{s:Results} we also show two plots based on the number of cases.

\subsection{\label{ss:Methods-EarlyStages} Early stages of a pandemic}

\noindent In the early stages when cases and deaths are still in their first rising phase, it is important to monitor the initial spread of the pandemic \citep{phua2020intensive, kennedy2020modeling}: How far behind are different countries compared to the country where the epidemic started and how fast is the current spread in each country? The most relevant quantities are the time lag with respect to the current epicenter of the pandemic and the doubling time.

\subsubsection{\label{sss:TimeLag} Time Lag}

%%%%%%%%%%%%%%%%%%%%%%%%%%%%%%%%%%%%%%%%%%%%%%%%%%
%%%%%%%%%%%%%%%%%%%%% Fig. 2 %%%%%%%%%%%%%%%%%%%%%
%%%%%%%%%%%%%%%%%%%%%%%%%%%%%%%%%%%%%%%%%%%%%%%%%%
%
\noindent In \textbf{Fig. \ref{Fig2_Time_Lags}} we show the course of the number of deaths for three selected countries (Italy, Spain and the US) during the initial period of the pandemic, from the first reported death in any of these three countries to the end of March. For better comparability the numbers were normalized to the overall population of the respective country. 

Italy was the country in the Western hemisphere with the earliest onset of an epidemic \citep{nacoti2020epicenter, indolfi2020outbreak} and also the first Western country with an officially recorded COVID-19 fatality which, as the graph shows, occurred on February 23, 2020, eight days earlier than in the US and eleven days earlier than Spain. In order to compare the course of the epidemic in different countries it thus makes sense to use Italy as the reference country. 

The first important information to know is by how many days your country lags behind Italy's curve or, at a later stage, after a potential reversal of fortune, by how many days it is ahead in its epidemic course. To this aim, we define the time lag of a country with respect to Italy as follows:

\begin{enumerate}
\item Start with the last data point of the country and then check when Italy's curve crossed this value.
\item Do vice versa in case Italy is behind. 
\end{enumerate}
%
% Figure 3
\begin{figure*}[!ht]
	\centering
	\includegraphics[width=1\textwidth]{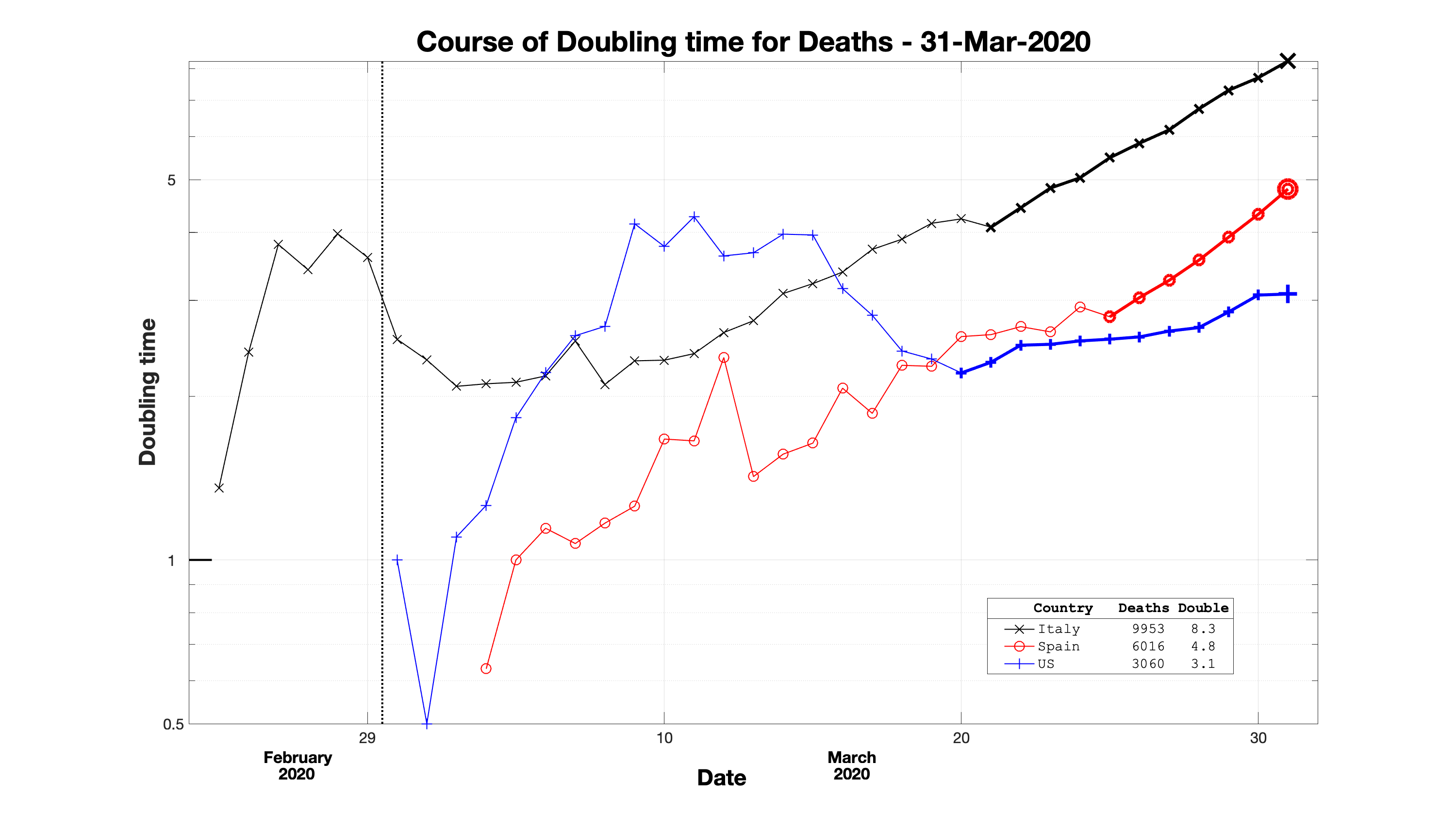}
	\caption{Doubling times of number of deaths for the same three countries and the same time interval used in Fig. \ref{Fig2_Time_Lags}. All three curves show quite large fluctuations early but are much smoother after that. At the end of March the doubling time was longest for Italy followed by Spain and the US. However, they were all trending towards larger values corresponding to a slowing down of the exponential growth in numbers of deaths.}
	\label{Fig3_Doubling_Times}
\end{figure*}

In Fig. \ref{Fig2_Time_Lags} this is illustrated for three different countries, using the end of March (marked as thick dashed vertical line 2) as an example. By construction the time lag of Italy to itself is always $0$. Spain reached $129$ deaths per one million inhabitants three days later than Italy. Accordingly, the time lag is $-3$. The US reached $9$ deaths per one million inhabitants $19$ days later than Italy. So here the time lag is $-19$.

Note that the time lag defined this way is a time-dependent variable. Instead of extracting via some kind of double fitting one value for the pair of two entire curves it is estimated such that it can change day by day. In our example from Fig. \ref{Fig2_Time_Lags} in the particular interval from March 11, 2020 (marked as thin vertical line 1) to March 31, 2020 (again line 2) the time lag between Spain and Italy decreased from $-9$ days to $-3$ days (Spain was catching up), while the time lag between the US and Italy increased from $-17$ days to $-19$ days (at that time the US was falling further behind).

\subsubsection{\label{sss:DoublingTime} Doubling time}

\noindent The second variable of importance is the doubling time, the characteristic unit for exponential growth. It is the time it takes for the number of deaths (or cases) to double. The lower its value the faster the spread of the epidemic. The first step in calculating this number is to determine the percentage growth rate $p(t)$ from one day to the next:  
\begin{equation} \label{eq:percentage_growth}
	p (t) = \frac{d(t)-d(t-1)}{d(t-1)}.
\end{equation}
%  percent_increase=y_data(t+1)-y_data(t)./y_data(t);
%
Here d(t) refers to the cumulative number of deaths until day t. From the percentage growth the doubling time $T_d$ is calculated as:
\begin{equation} \label{eq:doubling-time}
	T_d (t) = \frac{ln(2)}{ln[1 +p (t)]}.
\end{equation}
% doubling_time_mat=log(2)./log(1+percent_increase);

%%%%%%%%%%%%%%%%%%%%%%%%%%%%%%%%%%%%%%%%%%%%%%%%%%
%%%%%%%%%%%%%%%%%%%%% Fig. 3 %%%%%%%%%%%%%%%%%%%%%
%%%%%%%%%%%%%%%%%%%%%%%%%%%%%%%%%%%%%%%%%%%%%%%%%%
%
Like the time lag the doubling time is also a time-local quantity that changes day by day. In \textbf{Fig. \ref{Fig3_Doubling_Times}} we depict the course of the doubling times for the same three countries and during the same time interval that was already used in Fig. \ref{Fig2_Time_Lags}. In the beginning of the pandemic in each country the doubling time exhibits quite large fluctuations due to the low absolute numbers, but after a few weeks when the numbers rise this tends to stabilize into a more smooth course. 

Articles that have applied the doubling time in the context of the Covid19-pandemic include \cite{nunes2020visualising, lurie2020coronavirus}.
%
% Figure 4
\begin{figure*}[!ht]
	\centering
	\includegraphics[width=1\textwidth]{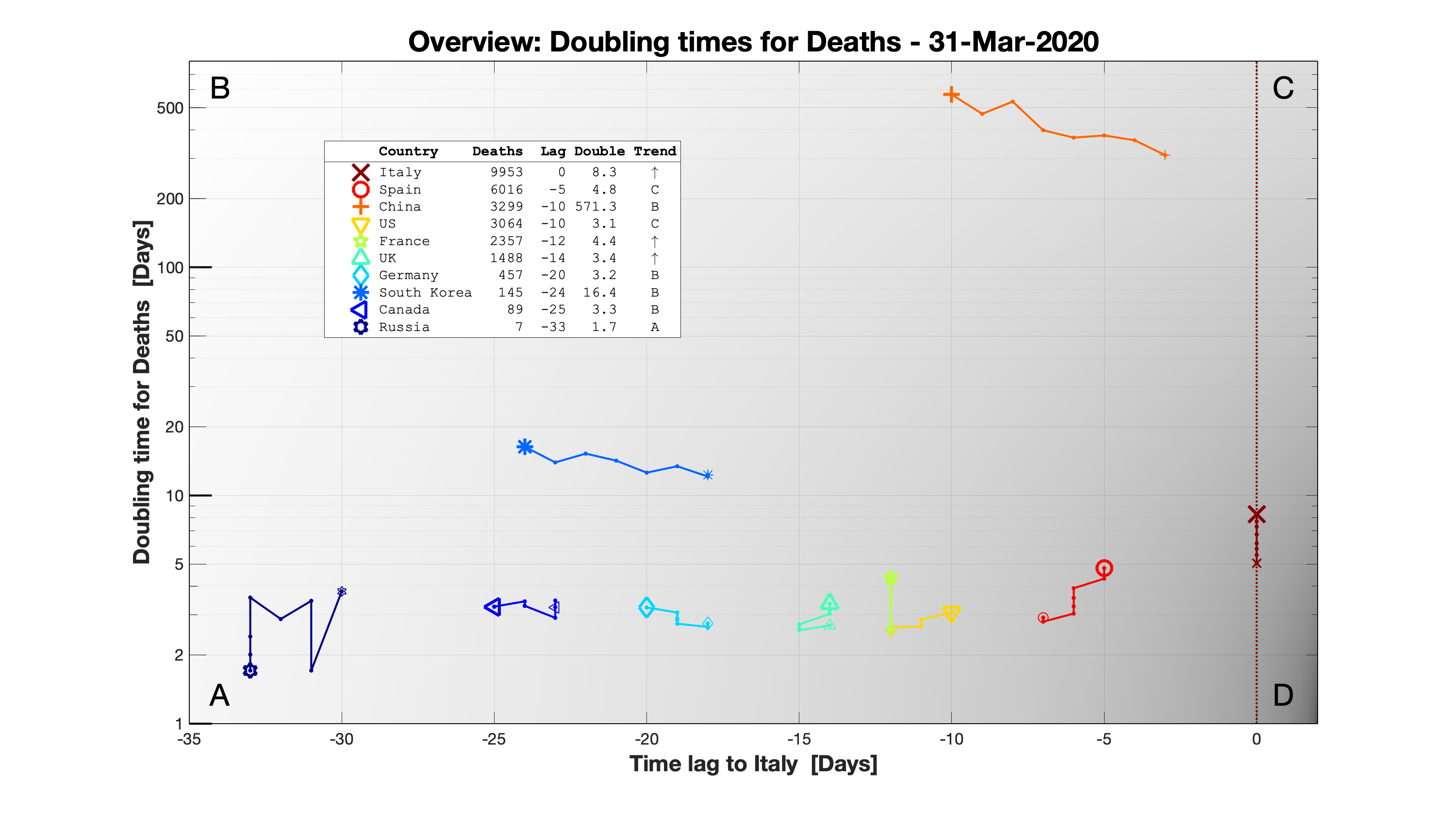}
	\caption{Doubling time versus time lag with respect to Italy for $10$ different countries. Large markers indicate the position on March 31, 2020, the tail shows the development over the previous seven days. The four corners are marked by letters A to D and the background is shaded to indicate more or less preferable regions (from bright to dark). The legend states overall number of deaths until the end of March 2020 as well as time lag, doubling time and trend over the last seven days.}
	\label{Fig4_4corners}
\end{figure*}

\subsubsection{\label{sss:Four-corners} Overview plots}

%%%%%%%%%%%%%%%%%%%%%%%%%%%%%%%%%%%%%%%%%%%%%%%%%%
%%%%%%%%%%%%%%%%%%%%% Fig. 4 %%%%%%%%%%%%%%%%%%%%%
%%%%%%%%%%%%%%%%%%%%%%%%%%%%%%%%%%%%%%%%%%%%%%%%%%
%
\noindent In the next step we combine these two time-local quantities, time lag and doubling time, in one large overview graph that allows for an easy comparison of the current state of the epidemic in many different countries. In \textbf{Fig. \ref{Fig4_4corners}} we plot the doubling time for deaths versus the time lag with respect to Italy for ten countries (including the ones from Figs. \ref{Fig2_Time_Lags} and \ref{Fig3_Doubling_Times}).

The countries that are compared with Italy were selected as follows: Four of them (Russia, South Korea, China and Spain) were at that moment in time closest to one of the four corners in the plot. The remaining five countries were on different stages of a rather typical curve of a country on this graph \citep{world2020coronavirus72}.
% (with some variability due to differences in choices made and measures taken or adhered to)

In this plot we use the brightness of the background to indicate the preferred order of the corners from Bright (B) to Dark (D): B, A, C, D. But our description begins with the typical starting point A:
%
% Figure 5
\begin{figure*}[!ht]
	\centering
	\includegraphics[width=1\textwidth]{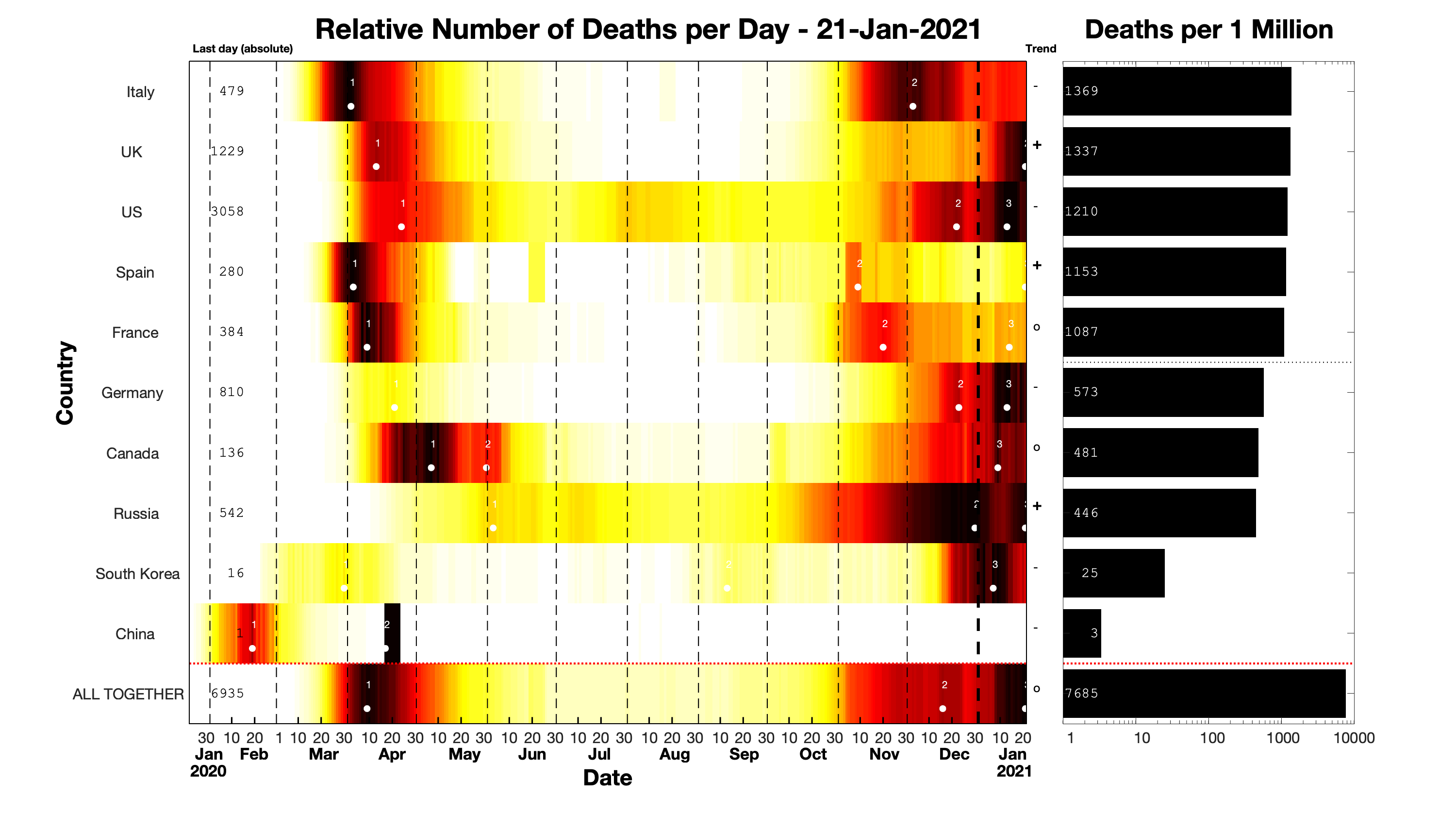}
	\caption{Number of reported new deaths per day for the ten countries from Fig. \ref{Fig4_4corners} and for the year from January 22, 2020 to January 21, 2021. Each row represents one country and colours are normalised from 0 (white) to 1 (black) by the maximum number of deaths of that country over the whole interval. The last row depicts the sum over all these countries together. Local maxima are marked by white bullet points and the ordinal number of the wave peak. On the left we indicate the absolute number of deaths on the last day and on the right the trend over the last week (`+',`-',`o' for upwards, downwards, constant, respectively). Finally, the histogram plot on the right shows for each country the number of deaths per one million inhabitants over the whole year. In order to provide information on the relative overall impact of the epidemic in different countries we here use this number as the criterion for the sorting of rows.
% Among the countries shown, these numbers were highest for Italy, the UK and the US.
}
	\label{Fig5_Wave_Detection}
\end{figure*}

\vspace{.4cm}

\noindent \textbf{A} – Large relative time lag, low doubling time.

\vspace{.1cm}

\noindent Basically all countries start on the lower left of the graph. There it is still early days, presumably the virus has started to spread within the country not that long ago so the time lag to the epicenter is typically quite big (depending on how long it took for the virus to reach the country, the later a country enters the plot the bigger the time lag). The initial doubling times are rather low so the spreading advances quickly but the numbers are still such that it might appear as if there is not yet that much to worry about. However, this is actually the time where measures should be taken as soon as possible in order to make a big difference later. At the end of March 2020 the closest country to this corner A was Russia which had just entered the plot with its first registered deaths.

\vspace{.4cm}

\noindent \textbf{B} – Large relative time lag, high doubling time.

\vspace{.1cm}

\noindent If a country is close to corner B it means that it has basically contained the virus in its earliest days and the doubling times are so high that one can hardly speak of an epidemic. On this day no country had really gotten there yet, but among the countries selected here South Korea was the one that was slowly getting closer.

\vspace{.4cm}

\noindent \textbf{C} – Relative time lag close to zero, high doubling time.

\vspace{.1cm}

\noindent This means that for the moment the worst is over but also that it was very bad. Eventually all countries tend to go up towards larger doubling times but of course it is much better to do it earlier rather than later. On the 31st of March 2020 China was the country closest to this corner but since less and less new deaths were reported it was actually moving towards corner B.

\vspace{.4cm}

\noindent \textbf{D} – Relative time lag close to zero, low doubling time.

\vspace{.1cm}

\noindent This is the situation to avoid at all costs (literally). Here countries are already right in the middle of an epidemic but the doubling times are still very low. This can be very bad because of the characteristics of unabated epidemic spread. At A it might take a few days to double the number of cases from $100$ to $200$ but at D it would take exactly the same time to double from $10.000$ to $20.000$ or even from $100.000$ to $200.000$ (depending on the overall stage of the epidemic). At the end of March 2020 the country closest to this situation was Spain (apart from the reference country Italy) and in fact it was right on its way of catching up with Italy.

\vspace{.4cm}

The remaining countries were at that point in time positioned somewhere between A and C. Like Spain, the US and the UK were moving closer to Italy. France was basically time-locked with Italy which means that its death curve was following Italy's with a constant time lag. By contrast, Germany and Canada were moving further away from Italy.

In Fig. \ref{Fig4_4corners} we show the position of all the countries in this two-dimensional plot at the end of March 2020 but to each country we also append a tail that depicts the development over the previous seven days. The direction of the movement over that week is captured in the trend which can be found as the very last entry in the legend:

\noindent \textbf{A} - towards lower doubling times but larger time lags with respect to Italy

\noindent $\leftarrow$ - no change in doubling time but towards larger time lags

\noindent \textbf{B} - towards higher doubling times and larger time lags (best possible course)

\noindent $\uparrow$ - towards higher doubling times, no change in time lag

\noindent \textbf{C} - towards higher doubling times but shorter time lags

\noindent $\rightarrow$ - no change in doubling time but towards shorter time lags

\noindent \textbf{D} - towards lower doubling times and shorter time lags (worst possible course)

\noindent $\downarrow$ - towards lower doubling times but no change in time lag

\noindent $\bullet$ - no change in either direction

Over this week, apart from Russia and Canada, the doubling time of most of these countries had increased. Regarding the time lag, there were three groups: for the countries on the right (Spain, the US and the UK) it had increased, for the countries on the left (Germany, South Korea, Canada and Russia) it had decreased and for the two countries in the middle (UK and France) it had remained constant. Accordingly, overall the trends were dominated by B, $\uparrow$ and C with only two countries on a downward trend towards A.

% #######################################################################################
% #######################################################################################
% #######################################################################################

\subsection{\label{ss:Methods-LaterStages} Later stages of a pandemic}

\noindent While the plots for the earlier stages are designed to provide a comparative overview of the initial rise, in the later stages the focus shifts to monitoring the course of the pandemic in each country in terms of peaks, valleys and plateaus in order to be able to react accordingly \citep{charpentier2020covid}. The transition between these two stages takes place around the time the daily numbers of cases and deaths stop rising for the first time \citep{leung2020first}.

%%%%%%%%%%%%%%%%%%%%%%%%%%%%%%%%%%%%%%%%%%%%%%%%%%
%%%%%%%%%%%%%%%%%%%%% Fig. 5 %%%%%%%%%%%%%%%%%%%%%
%%%%%%%%%%%%%%%%%%%%%%%%%%%%%%%%%%%%%%%%%%%%%%%%%%
%
The left side of \textbf{Fig. \ref{Fig5_Wave_Detection}} shows a 2D color plot of the number of new deaths per day for the same ten countries already depicted in Fig. \ref{Fig4_4corners} and for one year starting on January 22, 2020. Each row is normalized individually in order to provide an overview of the course of the epidemics for each country separately. This means that for each country the color scale ranges from zero daily deaths (white) to the maximum daily number of deaths over the whole interval (black). As a consequence the course of the pandemic even in countries with numbers of different orders of magnitude can be compared in the same plot. It also becomes immediately apparent whether a country is already over its peak and whether there are new waves; the brighter the colors on the last day, the further away a given country is from its peak value.

We also use the color plot as a wave detector by identifying for each country all the local maxima that fulfill the following two criteria: 

\vspace{.3cm}

\noindent - Minimum prominence $P_{min}$

\vspace{.1cm}

\noindent The prominence $P$ is defined as the smaller of the largest decrease in value on both side of the local maximum before encountering the next local maximum. For any given sequence of daily increases $D$ the largest possible prominence is $P_{max} = max(D) - min(D)$.

\vspace{.3cm}

\noindent - Minimum separation $S_{min}$ between consecutive local maxima

\vspace{.1cm}

\noindent The separation $S$ is defined in units of sample points (here days). For a given $S > 0$, we select the largest local maximum and ignore all other local maximum within $S$ units of it. This process is repeated until no more local maxima are detected.

\vspace{.3cm}

There is no unique and unambiguous definition of a wave peak, so varying these two parameter values will lead to different detections. Here we set the minimum prominence to $P_{min} = 0.5$ and the minimum separation to $S_{min} = 10$ days. This selection eliminates all minor bumps and maintains only the large-scale peaks that appear to be significant.

We also added trend indicators right next to the value of the current day that show whether the numbers from the last day are more than $5 \%$ higher than they were the week before (`+', upward trend), whether they are within $5 \%$ of that value (`o', plateau) or whether they are more than $5 \%$ lower (`-', downward trend).

The individual normalization used in the color plot facilitates tracking the course of the pandemic within each country and allows to infer relative time lags of peaks and valleys between different countries. However, it does not provide any information about the overall severity of the situation in each country. To rectify this we add a histogram (right) with the overall numbers for every country normalized by population size.
%
% Figure 6
\begin{figure*}[p]
	\centering	
	\begin{minipage}{1.02\textwidth}
	\includegraphics[width=1\textwidth]{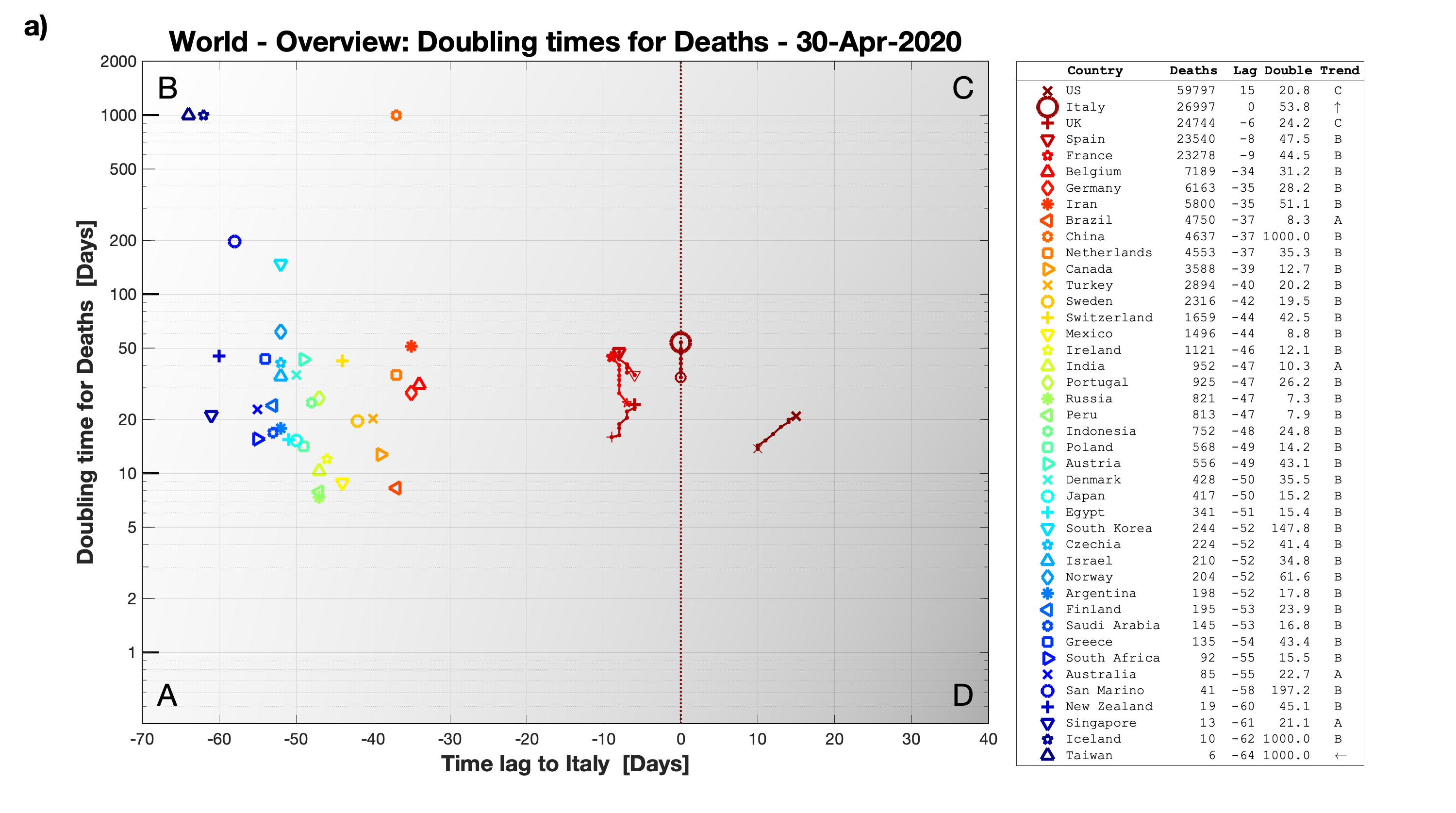}
	\includegraphics[width=1\textwidth]{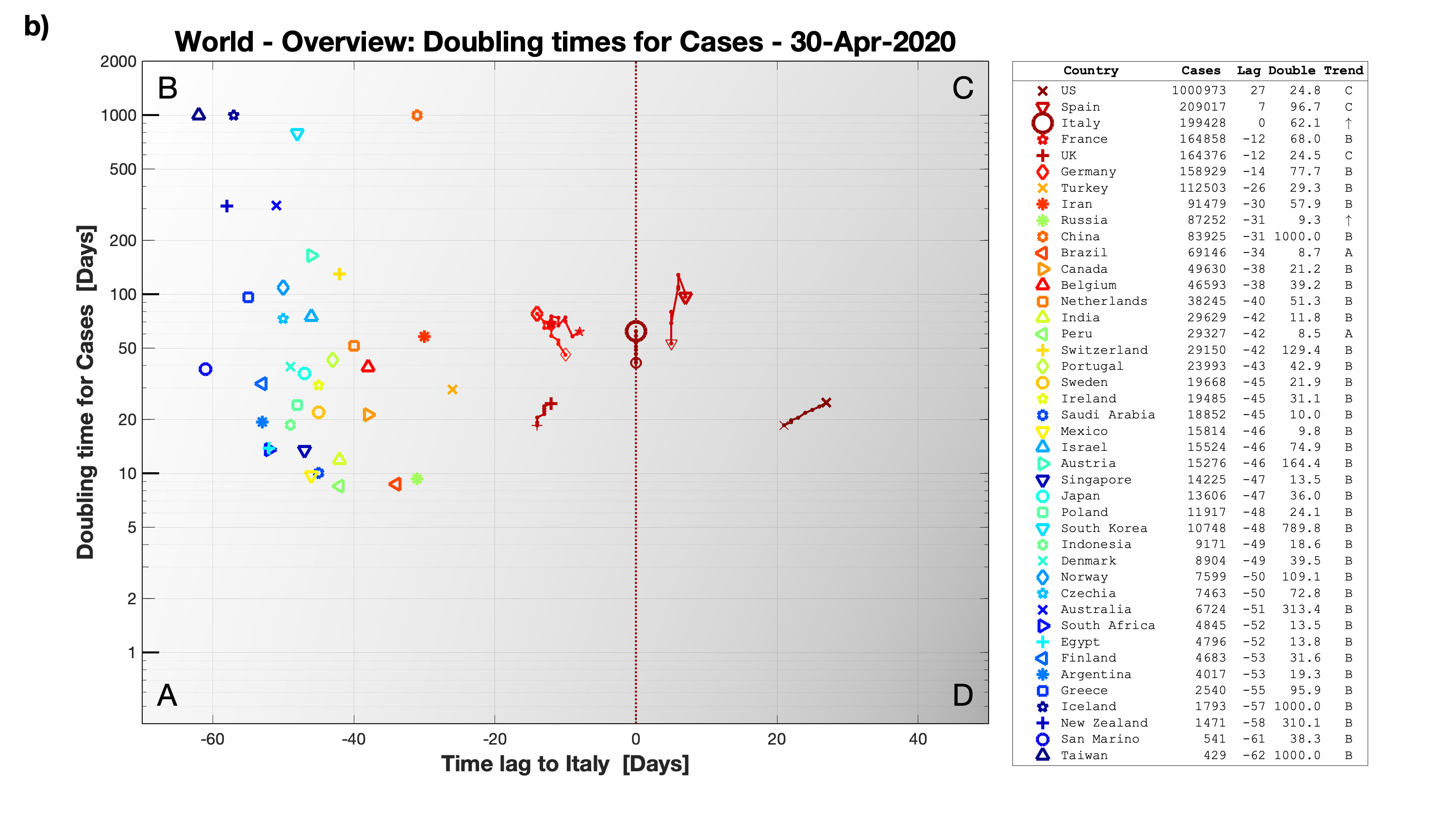}
	\end{minipage}
	\caption{Doubling time versus time lag with respect to Italy as in Fig. \ref{Fig4_4corners} but now for $42$ different countries and one month later, on April 30, 2020. For clarity, only the seven day tails of the eight countries that are most advanced in the pandemic (in terms of time lag) are shown, however, the seven day trends for all countries can be found in the legend. In deaths (a), at this point in time, the US had surpassed Italy as the worldwide leader of the pandemic, with the UK, Spain and France slightly behind, while for cases (b) not only the US but also Spain had surpassed Italy, and}
	\label{Fig6_4Corners-All}
\end{figure*}

Finally, depending on which information should be stressed, countries can be sorted in various ways:

\vspace{.3cm}

\noindent - The value of the histogram on the right hand side sorts countries by the overall severity of the situation (e.g., deaths per one million inhabitants). This sorting was used in Fig. \ref{Fig5_Wave_Detection} and will also be used in Figs. \ref{Fig7_Wave_Detection-All}a and \ref{Fig7_Wave_Detection-All}b for both worldwide deaths and worldwide cases.

\vspace{.3cm}

\noindent - The value of normalized new number of deaths/cases on the last day allows a comparison of the current state of the pandemic compared to the peak value of each country. Which countries are currently at their absolute peak and for which countries the worst is behind? This sorting will be used in Fig. \ref{Fig8_Wave_Detection-US} to compare the situation for all the US states.

\vspace{.3cm}

\noindent - The occurrence of the first death/case or the position of the peak of the first wave provides information about the gradual or sudden spatio-temporal propagation of the pandemic. Where did the pandemic start and where did it arrive last? In Fig. \ref{Fig9_Wave_Detection-Italy} we will use the sorting based on first cases to trace the initial spread of the virus in the Italian regions.

\vspace{.3cm}

\noindent - The similarity of the daily new deaths/cases profiles. We use a straightforward combination of correlation coefficient analysis and single linkage algorithm to cluster countries according to the similarity of their temporal profiles. From the resulting hierarchical dendrogram we obtain an order that starts with the countries that are most similar to each other and ends with those that are least similar to any of the other countries. This sorting will be used in Fig. \ref{Fig10_Correlations} for the worldwide data.

% ####################################################################################
% ####################################################################################
% ############################                          ##############################
% ############################          Results         ##############################
% ############################                          ##############################
% ####################################################################################
% ####################################################################################

\section{\label{s:Results}Results}

\subsection{\label{ss:Results-EarlyStages} Early stages of a pandemic}

\noindent In the early stages cases and deaths are still on their first rise and typically there is an early epicenter which becomes a very useful reference to which to compare the state of the pandemic in any given country. The most important indicator of this state is the doubling time. Thus, the two relevant quantities are the time lag with respect to that epicenter (here, Italy) and the doubling time and in \textbf{Fig. \ref{Fig6_4Corners-All}} we plot one against the other.

%%%%%%%%%%%%%%%%%%%%%%%%%%%%%%%%%%%%%%%%%%%%%%%%%%
%%%%%%%%%%%%%%%%%%%%% Fig. 6a %%%%%%%%%%%%%%%%%%%%
%%%%%%%%%%%%%%%%%%%%%%%%%%%%%%%%%%%%%%%%%%%%%%%%%%
%
First, in Fig. \ref{Fig6_4Corners-All}a we look at deaths numbers for all the $42$ countries on April 30, 2020 which was around the time the first worldwide peak in deaths had just passed \citep{world2020coronavirus102}. Fig. \ref{Fig6_4Corners-All}a is accompanied by \textbf{Supplementary Movie 1} which contains the development from the day Italy reported its second death up to the end of April (such that the final frame of the movie corresponds to this Fig. \ref{Fig6_4Corners-All}a).

At this moment in time countries could basically be divided into three different groups. The most severe situation was found in the US which actually had already surpassed Italy as the front runner of the pandemic and even at that advanced stage had a rather low doubling time of $20$ days and was thus continuing its course towards higher positive time lags. The other members of that group were the European countries Italy, the UK, Spain and France which were all more or less phase-locked with Italy, i.e. the time lags remained quite constant.

The second and by far largest group of countries was showing a trend towards increasing negative time lags and higher doubling times, thus getting closer to corner B (compare last entries in the legend). Typically this meant that for those countries the initial rise was already flattening considerably. This group included Iran, Germany and Belgium, which were all closest to Italy but still moving away.    

The third and last group consisted of Taiwan, Iceland, and China which had all basically stopped reporting any new deaths. At least for that specific moment in time these five countries had brought the pandemic under control.

%%%%%%%%%%%%%%%%%%%%%%%%%%%%%%%%%%%%%%%%%%%%%%%%%%
%%%%%%%%%%%%%%%%%%%%% Fig. 6b %%%%%%%%%%%%%%%%%%%%
%%%%%%%%%%%%%%%%%%%%%%%%%%%%%%%%%%%%%%%%%%%%%%%%%%
%
\textbf{Fig. \ref{Fig6_4Corners-All}b} depicts the same kind of plot as Fig. \ref{Fig6_4Corners-All}a, but now for cases instead of deaths. As before, \textbf{Supplementary Movie 2} shows the whole history of this plot from the beginning of February (first reported cases in Italy) to the end of April. Overall, both the groupings and trends for cases are very similar to the ones seen for deaths. One notable difference is that for cases the separation between the three groups is much less pronounced.

% #######################################################################################
% #######################################################################################
% #######################################################################################

\subsection{\label{ss:Results-LaterStages} Later stages of a pandemic}

% Figure 7
\begin{figure*}[p]
	\centering
	\begin{minipage}{1.05\textwidth}
	\includegraphics[width=1\textwidth]{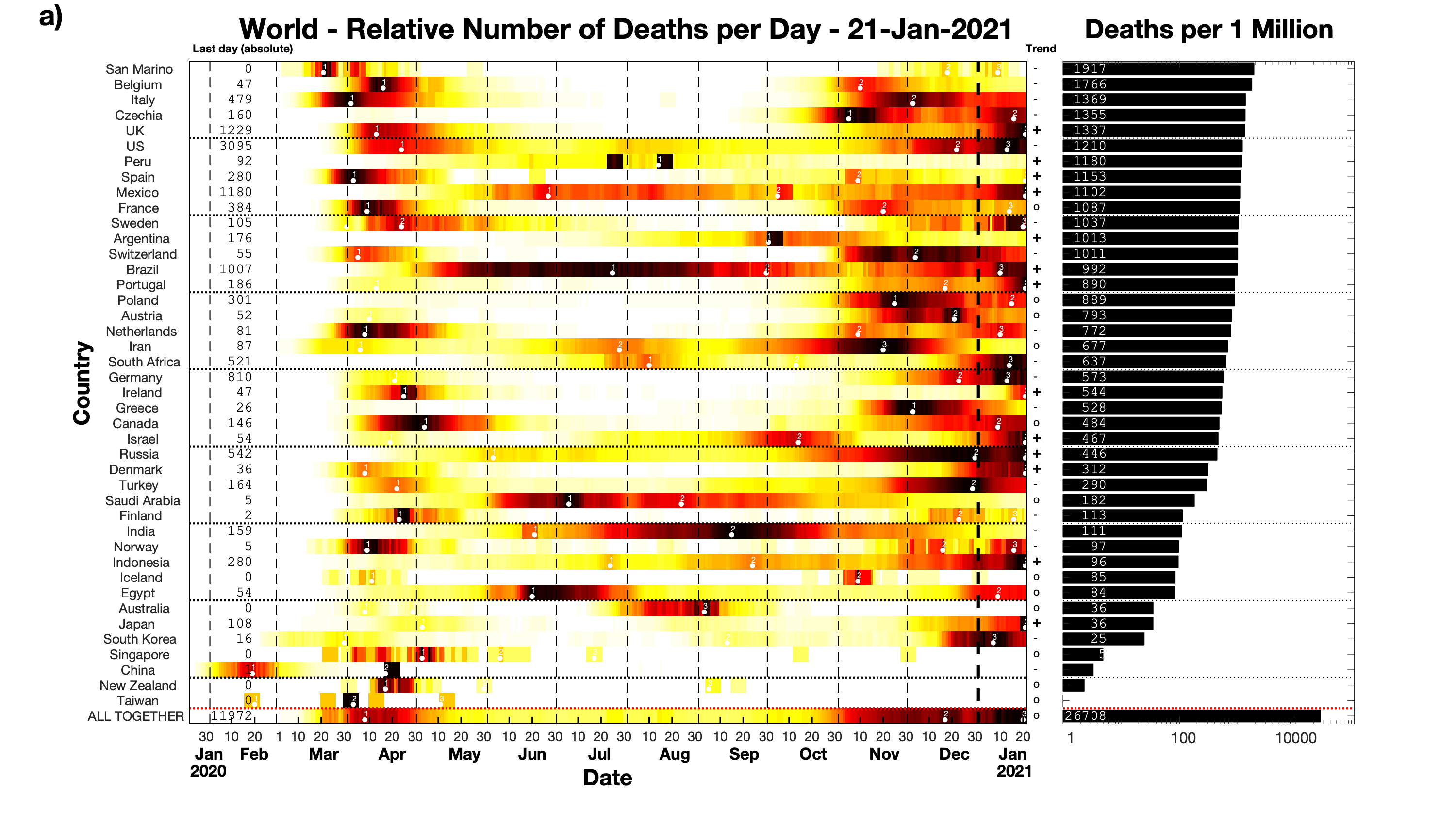}
	\includegraphics[width=1\textwidth]{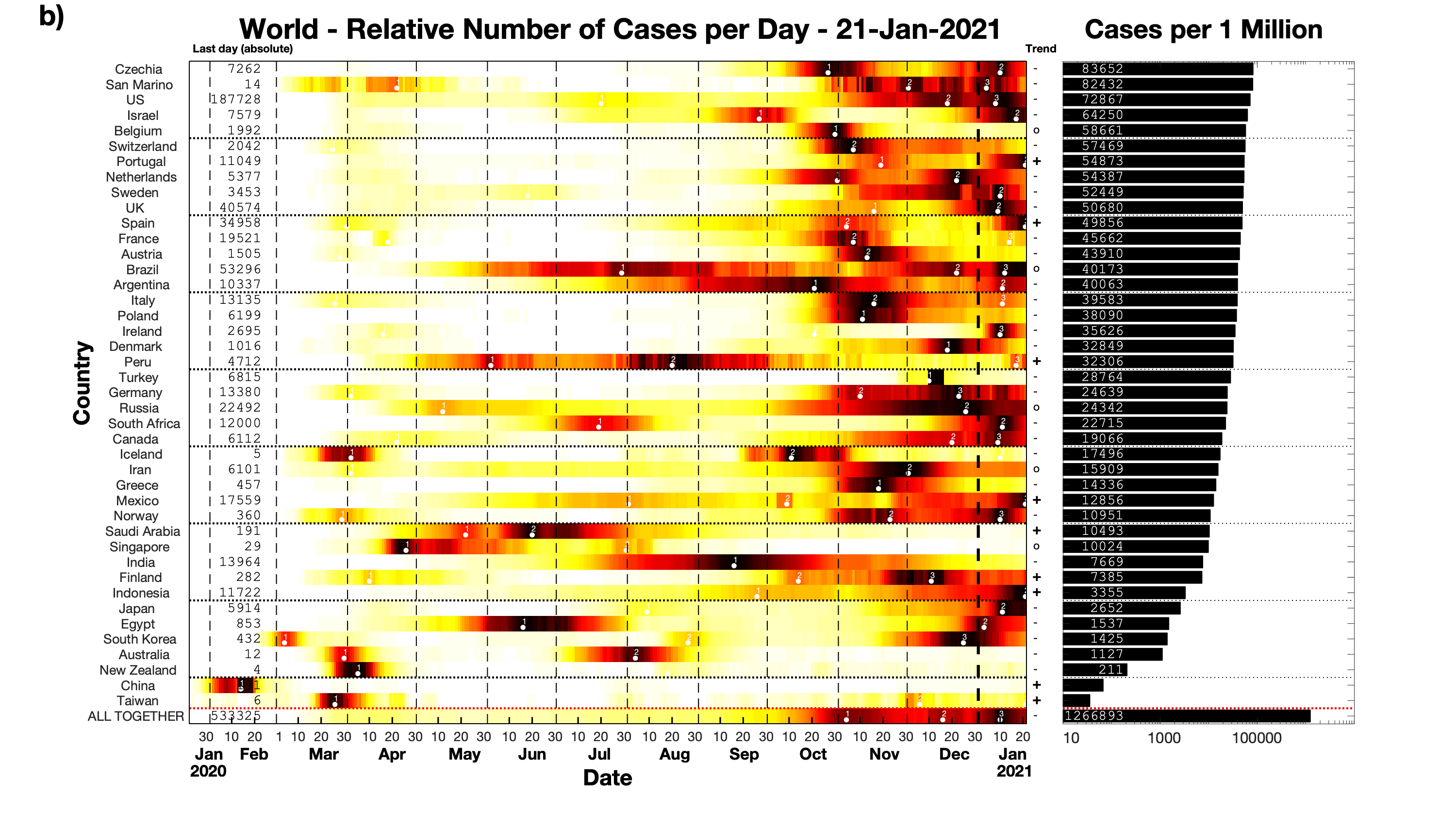}
	\end{minipage}
	\caption{Number of reported new deaths (a) and cases (b) per day for the year from January 22, 2020 to January 21, 2021 for all $42$ selected countries. Layout as in Fig. \ref{Fig5_Wave_Detection}. Countries are again sorted by the number of deaths per one million inhabitants over the whole year (histogram on the right). At this moment in time more than half of the countries were peaking or close to peaking and typically it was either the second or the third peak.}
	\label{Fig7_Wave_Detection-All}
\end{figure*}
%
% Figures 8 and 9 (on one page)
\begin{figure*}[p]
	\centering
	\includegraphics[width=1.02\textwidth]{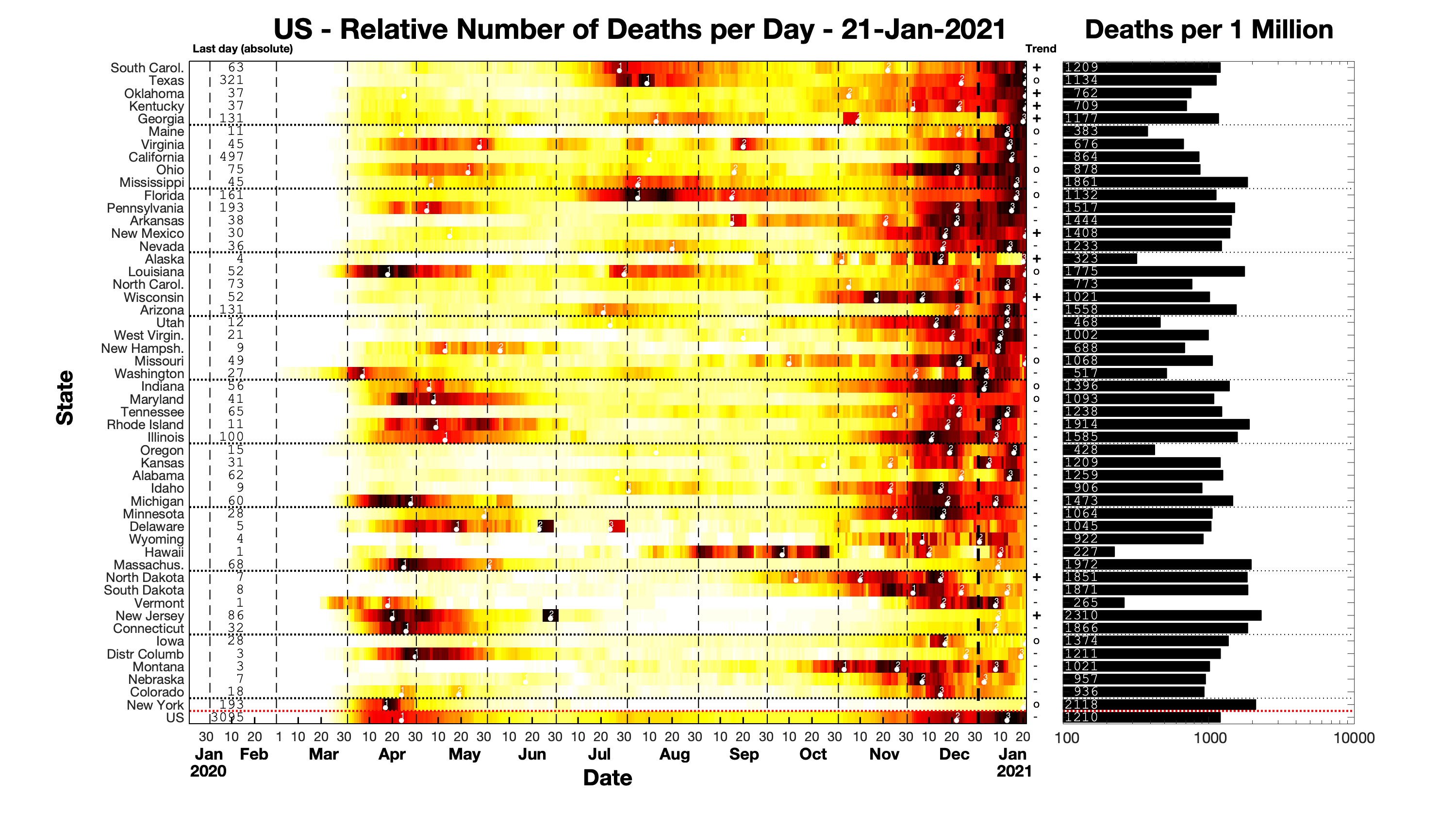}
	\caption{Number of reported new deaths per day for all the US states during the year from January 22, 2020 to January 21, 2021. Layout as in Fig. \ref{Fig5_Wave_Detection} but in contrast to the previous wave detection plots (Figs. \ref{Fig5_Wave_Detection} and \ref{Fig7_Wave_Detection-All}) this one is sorted according to the (normalized) value obtained for the very last day of the plot. This provides information about the relative severity of the situation in each state at this point in time, while the histogram on the right still shows the overall severity up to that day.}
	\label{Fig8_Wave_Detection-US}
	\vspace{-.15cm}
	%\vspace{.5cm}  % ######### use again ######### width was 1.03 (both here and above)
	\includegraphics[width=0.94\textwidth]{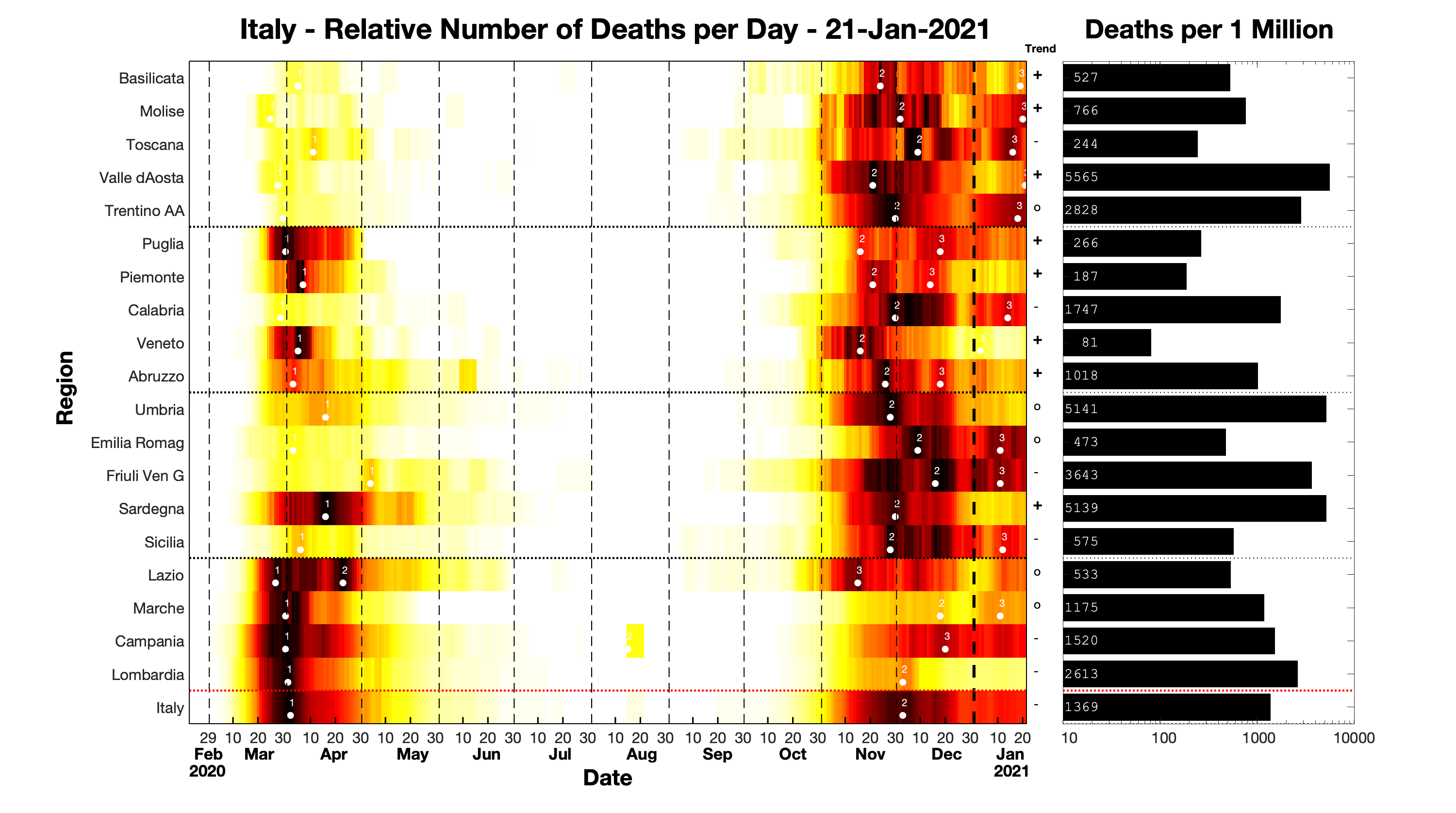}
	\caption{Number of reported new deaths per day for the Italian regions during the same year as before. Layout as in Fig. \ref{Fig5_Wave_Detection} with the only difference that in this wave detection plot the regions are sorted according to the time of the first reported death (from bottom to top). This allows tracing the course of the pandemic in Italy from early epicenters such as Lombardy and Campania to regions like Basilicata that were reached only much later.}
	\label{Fig9_Wave_Detection-Italy}
\end{figure*}

%
% Figure 10
\begin{figure*}[p]
	\begin{minipage}{1.03\textwidth}
		\includegraphics[width=1\textwidth]{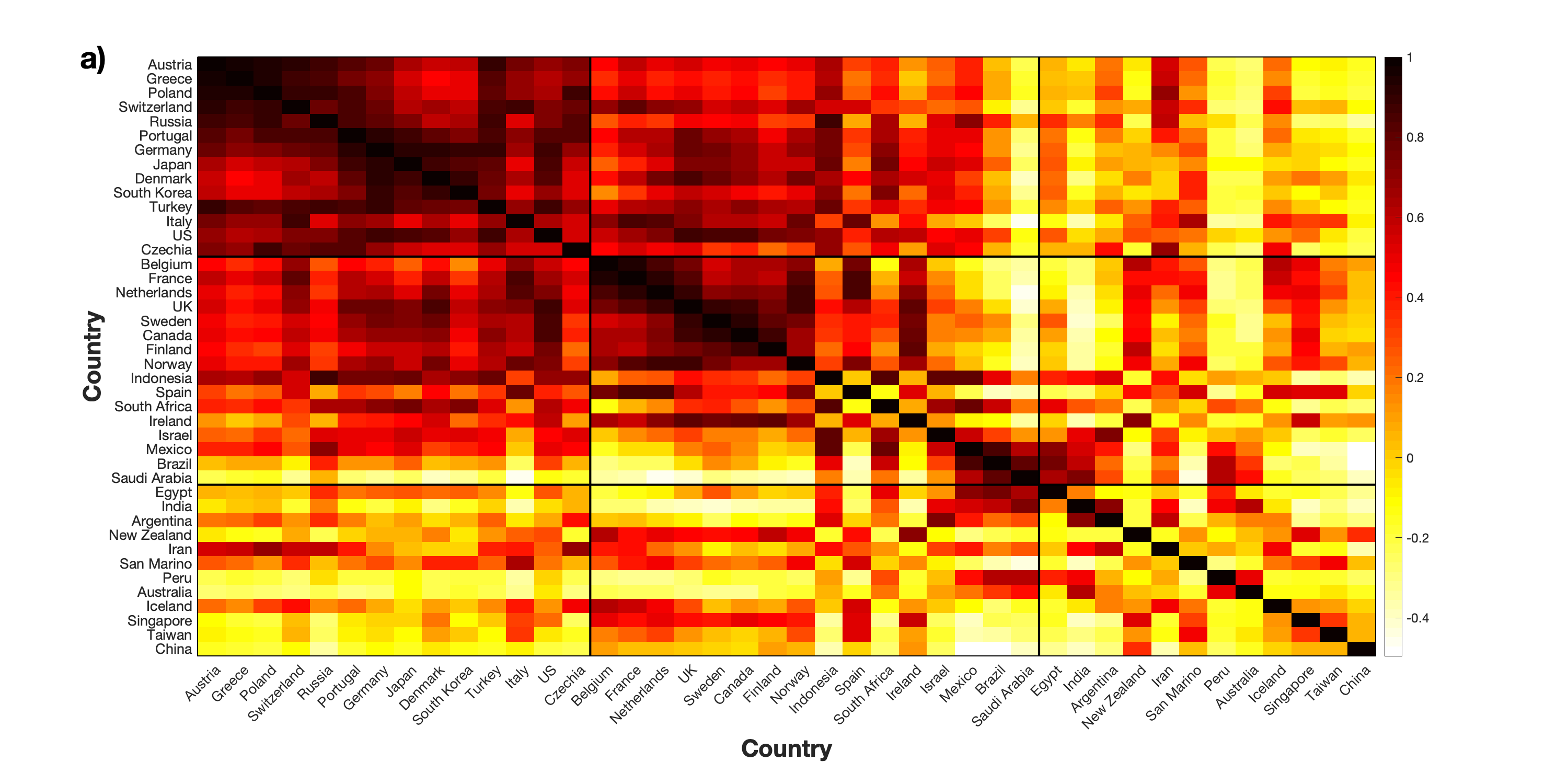}
		\includegraphics[width=1\textwidth]{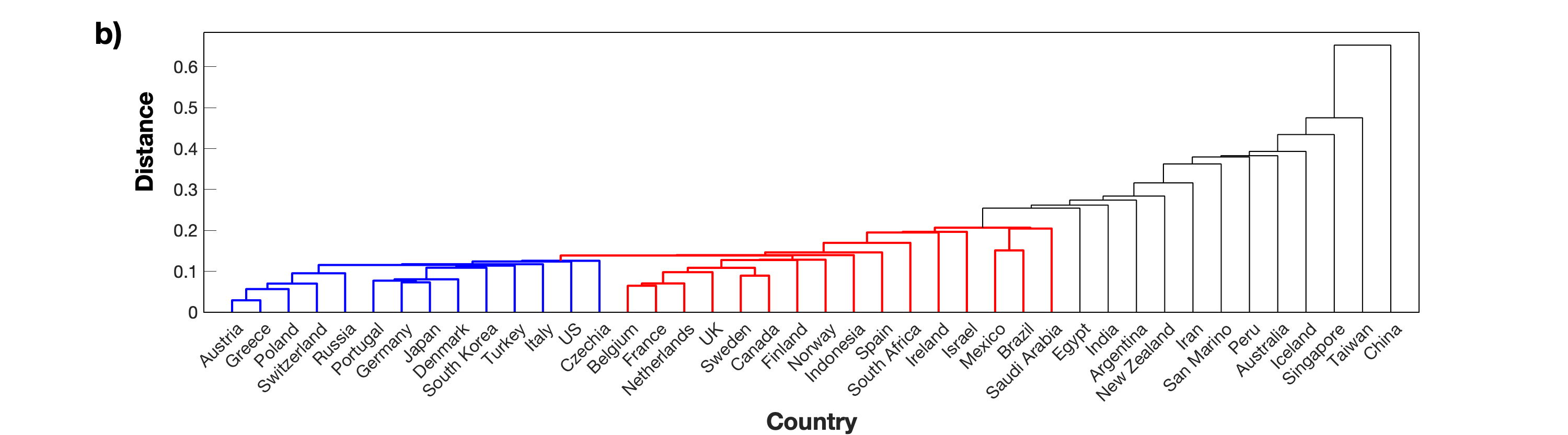}
		\vspace{-.3cm}
		\includegraphics[width=1\textwidth]{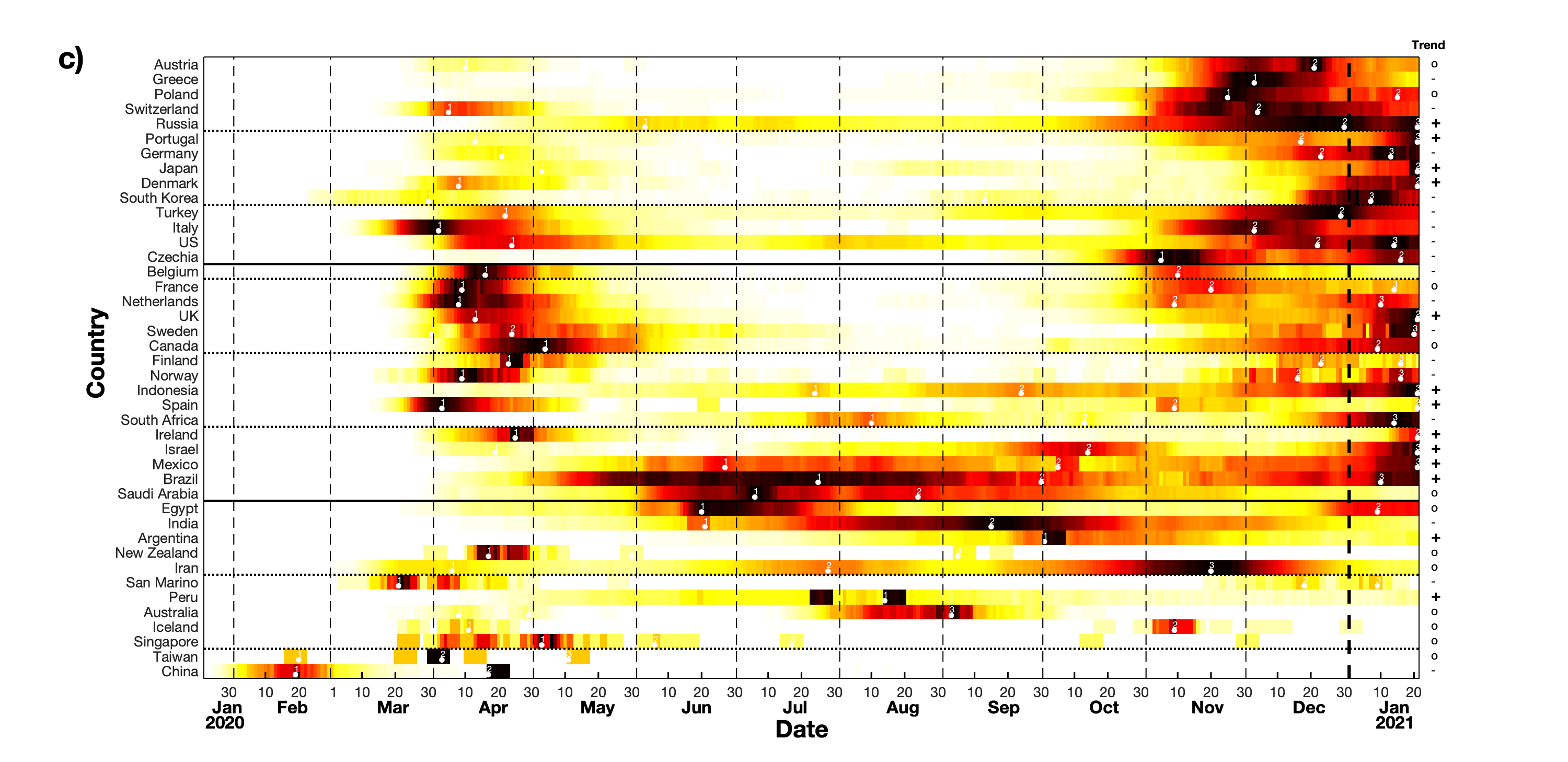}
	\end{minipage}
	\caption{Similarity of countries in terms of daily new deaths. (a) Pairwise Correlation coefficient matrix calculated from the temporal profiles as depicted in Fig. \ref{Fig7_Wave_Detection-All}a. (b) Hierarchical cluster tree (dendrogram) obtained from (a) via the single linkage clustering algorithm. Countries that are most similar to each other (lowest distances between them) are on the left, countries that are more unique (higher distances) are on the right. (c) Same data as in the wave plot of Fig. \ref{Fig7_Wave_Detection-All}a but with the countries sorted according to the similarity criterion. For clarity the same sorting was already used in (a). Note that the three colors in (b) and the corresponding separating lines in (a) and (c) are just visual aids, they are not based on any strict criteria.}
	\label{Fig10_Correlations}
\end{figure*}

%%%%%%%%%%%%%%%%%%%%%%%%%%%%%%%%%%%%%%%%%%%%%%%%%%
%%%%%%%%%%%%%%%%%%%%% Fig. 7a %%%%%%%%%%%%%%%%%%%%
%%%%%%%%%%%%%%%%%%%%%%%%%%%%%%%%%%%%%%%%%%%%%%%%%%
%
\noindent Once the initial rising phase of a pandemic has been passed, it becomes more important to monitor the rise and fall of deaths and cases via the characteristic peaks, valleys and plateaus in the profile of each country. Accordingly, we continue with \textbf{Fig. \ref{Fig7_Wave_Detection-All}a} which follows Fig. \ref{Fig5_Wave_Detection} in showing the number of reported daily new deaths from the beginning of the dataset up to a year later, but it does so for all $42$ countries. Similar to before, Fig. \ref{Fig7_Wave_Detection-All}a is the last frame of \textbf{Supplementary Movie 3}, which tracks the development of the number of deaths over the whole year.

After this one year, the countries hit the hardest were predominantly in Europe, but also countries like the US, Peru and Mexico had suffered very high numbers of deaths. On the other hand, many countries in Asia and Oceania had fared quite will. Among these countries were Taiwan, South Korea, Japan as well as New Zealand and Australia.

At this specific point in time there was almost equal division among upwards, constant and downwards trends, but there were just a few more countries trending downwards rather than upwards. So most countries were either plateauing on an unprecedented high level or had just started to slightly decrease (for confirmation just refer to the data for all countries together in the last row). The relatively highest peaks on that day were obtained for the US and Mexico, whereas a few countries (Taiwan, New Zealand, Singapore, Australia and Iceland) were reporting no deaths at all.

%%%%%%%%%%%%%%%%%%%%%%%%%%%%%%%%%%%%%%%%%%%%%%%%%%%
%%%%%%%%%%%%%%%%%%%%% Fig. 7b %%%%%%%%%%%%%%%%%%%%%
%%%%%%%%%%%%%%%%%%%%%%%%%%%%%%%%%%%%%%%%%%%%%%%%%%%
%
\textbf{Fig. \ref{Fig7_Wave_Detection-All}b} shows the reported number of daily cases for the same countries and the same time interval. The overall situation was quite similar to the one reported for deaths (Fig. \ref{Fig7_Wave_Detection-All}a). Also here half the countries were peaking or close to peaking and again only a handful of countries (Taiwan, China, New Zealand, Australia, and Iceland) reported almost no new cases at all. Fig. \ref{Fig7_Wave_Detection-All}b corresponds to the last frame of the final \textbf{Supplementary Movie 4}.

When comparing the relative positions of the countries in the two graphs for deaths (Fig. \ref{Fig7_Wave_Detection-All}a) and cases (Fig. \ref{Fig7_Wave_Detection-All}b) we find a rather high correlation coefficient of $0.84$, which means that typically countries that rank high in deaths also rank high in cases. The two most notable exceptions are Israel (which ranked much higher in cases than in deaths) and Mexico (the other way around, it ranked much higher in deaths than in cases).

So far we looked at global plots containing a comparison of different countries from all over the world. The last two plots depict the same kind of data but now for smaller regions within a country - first the US states and then the regions of Italy - and each time using a new kind of sorting criterion.

%%%%%%%%%%%%%%%%%%%%%%%%%%%%%%%%%%%%%%%%%%%%%%%%%%%
%%%%%%%%%%%%%%%%%%%%% Fig. 8 %%%%%%%%%%%%%%%%%%%%%%
%%%%%%%%%%%%%%%%%%%%%%%%%%%%%%%%%%%%%%%%%%%%%%%%%%%
%
First, in \textbf{Fig. \ref{Fig8_Wave_Detection-US}} we visualize the data of the US states (plus the country as a whole), again over the same first year of data availability. In this graph we sort the states according to the number of deaths on the very last day (in this case January 21, 2021) normalized to the maximum value obtained for each state so far. This gives us a good idea of the relative severity of the situation on that day since it sorts states according to how close countries are to their absolute peak. The states that are currently peaking can be found at the top whereas the states that have passed their peak(s) appear at the bottom.

On this very day in the US mostly southern states like South Carolina, Oklahoma, Kentucky, Texas, and Georgia were still peaking in deaths, while the state furthest away from its own past peak was New York (followed by Colorado and Nebraska). On the other hand, as the histogram on the right shows, the highest overall numbers of deaths had been obtained in north-eastern states such as New Jersey, New York, Massachusetts and Rhode Island.

\textbf{Fig. \ref{Fig9_Wave_Detection-Italy}} depicts the regions of Italy sorted by the time of their first reported death. It started with Lombardy in the North which became the early epicenter and then gradually reached the whole country with Basilicata being the last region to report a death.

This plot also shows quite nicely how the severity of the situation in each region can vary for different waves. Marche and Lombardia, two of the regions that were hit early and hard during the first wave (in fact were among the first places in Europe to report Corona-related deaths), were relatively speaking spared during the the second wave of the next winter, 2020/2021. On the other hand, regions such as Basilicata, Molise and Toscana had much worse trajectories the second time around.  

%%%%%%%%%%%%%%%%%%%%%%%%%%%%%%%%%%%%%%%%%%%%%%%%%%%
%%%%%%%%%%%%%%%%%%%%% Fig. 10 %%%%%%%%%%%%%%%%%%%%%
%%%%%%%%%%%%%%%%%%%%%%%%%%%%%%%%%%%%%%%%%%%%%%%%%%%
%
Finally, in \textbf{Fig. \ref{Fig10_Correlations}} we return to the worldwide data and 
combine a correlation coefficient analysis (Fig. \ref{Fig10_Correlations}a) with the single linkage algorithm to arrange the countries in a hierarchical cluster tree (dendrogram, Fig. \ref{Fig10_Correlations}b) according to the similarity of their daily new deaths profiles (from Fig. \ref{Fig7_Wave_Detection-All}a). Note that during this mapping from the pairwise distance matrix to a one-dimensional order important information gets lost. Vicinity in the ordered list does not correspond to a direct measure of distance. However, what hold is that countries that are most similar to each other appear on the left and countries with the most unique profiles can be found on the right (as reflected my the monotonous increase of the distances from left to right).

Using the resulting order from the dendrogram we get a new wave plot (Fig. \ref{Fig10_Correlations}c). Now we have on top the countries closest to each other which seem to be those with a rather weak first wave (spring 2020) but a very strong second wave (winter 2020/21). The next group contains the countries with two rather strong waves. Both of these groups are predominantly European and North American. They are followed by countries with a wave in summer 2020. This might be due to a later arrival time of the virus but there are also many countries from the Southern hemisphere which also points to an explanation in terms of anti-phase seasonal variations between the two hemispheres \citep{merow2020seasonality}. The last group include those that apart from a few sporadic eruptions had brought the pandemic mostly under control and ends which China with its very early outbreak followed by a consistent flattening of the curve.

% ####################################################################################
% ####################################################################################
% ############################                          ##############################
% ############################        Discussion        ##############################
% ############################                          ##############################
% ####################################################################################
% ####################################################################################

\FloatBarrier

\section{\label{s:Discussion}Discussion}

% Summary
\noindent We presented visualizations of epidemiological data (such as the number of deaths and cases) that take into account many countries (states, regions, etc.) at the same time and are tailored to the specific stage of the epidemic. During the initial rising phase we focus on time lag with respect to the current epicenter and doubling time since this combination can inform about the timeliness and the urgency of interventions needed to curb the spread. In contrast, during the later stages we monitor the state of the pandemic by following the wavelike profile of daily new death or case numbers with regard to peaks, valleys and plateaus. This, together with other epidemiological quantities such as attack rate \citep{liu2020secondary} or basic reproduction number \citep{park2020reconciling, viceconte2020covid, tao2020maximum}, can help to decide about appropriate counter measures, i.e. whether to impose, tighten or relax contact restrictions in the population \citep{anderson2020will, valba2020self}. % incidence

% Advantages, Methodological improvements and possible extensions
The visualizations used here are universal and can easily be applied to other kinds of epidemiological data. The spatial scale is flexible as well. While we focused on countries, states and regions, it would of course also work with smaller areas. Moreover, in the two-dimensional plots designed for the initial stages of an epidemic (Figs. \ref{Fig4_4corners} and \ref{Fig6_4Corners-All}) we use a country of reference, namely Italy, the early European epicenter of the pandemic \citep{nacoti2020epicenter, indolfi2020outbreak}. However, this is certainly a matter of choice. Different countries could be chosen, e.g., in order to test different hypotheses. Or you could select your own country and then the plot can basically be seen from your countries' point of view and provide more detailed information about its relative position in the pandemic. Finally, our methodology based on time lags and doubling times was only applied to the rising phase of the first wave (and indeed there it is most useful), but of course in cases where the individual waves are separate enough (e.g., due to seasonal variations \citep{merow2020seasonality}) it would also be possible to look at second, third or even later waves.

The color-coded wave detection plots used during the later stages can easily be modified to be sensitive to several other traits in the data. Here we used four different sorting criterions (Figs. \ref{Fig7_Wave_Detection-All}, \ref{Fig8_Wave_Detection-US}, \ref{Fig9_Wave_Detection-Italy} and \ref{Fig10_Correlations}c) but also the normalization of the color scale could be altered to stress certain other aspects of the data. Similarly, one could adapt the wave detection parameters in order to focus on specific time scales and resolutions.

% Comparison with existing methods ?????
% Disclaimers, Caveats (Data Quality), Outlook
Note that the current study itself is not concerned with drawing specific conclusions from the data, rather we focus on efficient and informative ways of presenting the data to then be in a better position to actually draw specific conclusions. We restrict ourselves to examinations of the past up to the present day, but there is no extrapolation into the future based on any kind of model, assumption or parameter selections. This way we avoid any pitfalls caused by potential deficiencies in either completeness or accuracy of the data \citep{sridhar2020modelling}. However, given reliable data these visualizations could be used to, e.g., correlate the data with different containment strategies \citep{lurie2020coronavirus}, or to perform other more extended analyses \citep{callaghan2020covid, edwards2021now}.

We would like to close with the following appeal: Please always be aware of the tragic real-life consequences behind these numbers and do whatever you can to keep them low or bring them down again.

% ####################################################################################
% ####################################################################################
% ####################################################################################

\section{\label{s:Outreach} Source codes and outreach}

\noindent Matlab source codes will be available at this webpage:

\vspace{.2cm}

\noindent \url{http://wwwold.fi.isc.cnr.it/users/thomas.kreuz/Source-Code/Corona.html}
% ZZZZZ under construction, not yet active, update zip-file and PDF of paper ZZZZZ

\vspace{.2cm}

\noindent Since the worsening of the pandemic in Italy (March 2020) regular updates based on the data analysis methods described in this article have been posted on this Facebook page:

\vspace{.2cm}

\noindent \url{https://www.facebook.com/tk.corona.updates/}

\vspace{.2cm}

\noindent This will continue as long as the pandemic causes significant damage all over the world. Regular updates show data for the same $42$ selected countries used here. On demand the data from the regions/provinces of some individual countries with publicly available data (specifically the US, Italy, Germany, Spain, the UK, and Canada) are displayed as well. For now this is mostly the US, the only major country with continuously elevated numbers and strong regional diversity.

% ####################################################################################
% ####################################################################################
% ####################################################################################

\section{\label{s:SupplementaryMovies} Supplementary Movies}

\noindent Caption \textbf{Supplementary Movie 1} 

\noindent (\url{https://youtu.be/2BMZw5PHmK0}):
% Corona_Movie_6.avi, 70 frames

\vspace{.3cm}

\noindent The first supplementary movie depicts the doubling time for deaths versus the time lag with respect to Italy for all $42$ countries from February 21, 2020 (the day Italy reported its second death) to April 30, 2020. The layout is identical to the one used in Fig. \ref{Fig6_4Corners-All}a, in fact, it ends with Fig. \ref{Fig6_4Corners-All}, the state of the pandemic at the end of April 2020. Over the course of time most countries tend to slowly move towards corner B (larger time lags with respect to Italy, higher doubling times). But there are a few exceptions, notably Spain, France, the UK, and in particular the US which during April can be seen to slowly overtake Italy.

\vspace{.5cm}

\noindent Caption \textbf{Supplementary Movie 2} 

\noindent (\url{https://youtu.be/vAsSQfWvJwQ}):
% Corona_Movie_7.avi, 90 frames

\vspace{.3cm}

\noindent The second supplementary movie is similar to the first but depicts cases instead of deaths. It runs from February 2, 2020 (the day Italy reported its second case) to April 30, 2020. Its last frame is identical to Fig. \ref{Fig6_4Corners-All}b.

\vspace{.5cm}

\noindent Caption \textbf{Supplementary Movie 3} 

\noindent (\url{https://youtu.be/UC13sP7gHeU}): % (Corona 3)
% Corona_Movie_8.avi, 365 frames

\vspace{.3cm}

\noindent The third supplementary movie shows the course of the pandemic for all $42$ countries over one year, from January 22, 2020 to January 21, 2021, in terms of number of reported new deaths per day. The layout is identical to the one used in Fig. \ref{Fig7_Wave_Detection-All}a. Countries are again sorted by the relative number of deaths over the whole year (histogram on the right) which makes it easier to follow the overall impact of the epidemic for different regions. During the initial stages China, Italy, South Korea, Iran and Spain were the epicenters of the pandemic, at the later stages Belgium, Czechia, Peru, the UK and the US were among those countries that exhibited the highest relative numbers of deaths. 

\vspace{.5cm}

\noindent Caption \textbf{Supplementary Movie 4}

\noindent (\url{https://youtu.be/WCyhIEQJc7Q}): % (Corona 4)
% Corona_Movie_9.avi, 365 frames

\vspace{.3cm}

\noindent The fourth supplementary movie is similar to the third but instead of deaths it depicts cases for all $42$ countries over one year. Its last frame corresponds to Fig. \ref{Fig7_Wave_Detection-All}b. This movie shows even more clearly how the pandemic spread from China and its neighboring countries all over the world.

\vspace{.5cm}

\noindent Further movies (various combinations of deaths/cases with/without normalization and different kinds of sortings for the $42$ selected countries, the US states, and the Italian regions can be found on the tk.corona.updates Youtube channel: 

\vspace{.2cm}

\noindent \url{https://www.youtube.com/channel/UCASTaaV9CKZEpsFRhnYF-EQ}

\vspace{.2cm}

\noindent The first four movies on the channel correspond to the Supplementary Movies 1-4 of this article.

\begin{acknowledgments}
T.K. would like to thank Alban Levy for useful discussions, his thorough reading of the manuscript as well as his very early efforts in rising awareness about the upcoming pandemic. T.K. would also like to thank Don MacLeod, Kaare Bjarke Mikkelsen, and Sabine Raphael for feedback on the data analysis and Benedetta Moschitta for many inspiring discussions.
\end{acknowledgments}

\bibliography{Kreuz_Corona_Bibliography}

\end{document}